\documentclass[nofootinbib,twocolumn,secnumarabic,amssymb, %nobibnotes,
aps,pra]{revtex4-2}
\usepackage{xcolor,soul,graphicx,amsmath,amssymb,epstopdf,amsthm,mathrsfs,latexsym,xfrac,enumerate,extarrows,titlesec,setspace,hyperref,xfrac,float,tabularx,subcaption%,biblatex
,cleveref}
\usepackage{color}
\usepackage[utf8]{inputenc}
\usepackage[english]{babel}
\usepackage{ulem}

\newcommand{\hide}[1]{}

\newcommand{\eq}[1]{Eq.\,(\ref{#1})}

\newcommand{\fig}[1]{Fig.\,\ref{#1}}

\newcommand{\nofig}[1]{\ref{#1}}

\begin{document}

\title{Superradiance-assisted two-color Doppler cooling of molecules}

\author{Caleb Heuvel-Horwitz}
\affiliation{Harvard Physics Dept, 17 Oxford St, Cambridge, MA 02138}
\author{S. F. Yelin}
\affiliation{Harvard Physics Dept, 17 Oxford St, Cambridge, MA 02138}
\date{\today}

\begin{abstract}
    For experiments that require a quantum system to be in the ultra-cold regime, laser cooling is an essential tool. While techniques for laser cooling ions and neutral atoms have been refined and temperatures below the Doppler limit have been achieved, present-day techniques are limited to a small class of molecules. %A cooling method does exist for diatomic molecules, but currently the only proposal for laser cooling of polyatomic molecules of greater than six atoms does not achieve temperatures desirable for most practical uses. 
    This paper proposes a general cooling scheme for  molecules based on vibrational-state transitions. %, using an effective four-level scheme with multiple electronic ground states, that is, vibrational and rotational sub-levels. 
    Superradiance is used to speed up the two-photon transition. Simulations of this scheme achieve temperatures comparable to those achieved by existing two-level schemes for neutral atoms and ions. 
\end{abstract}

\maketitle

\section{Introduction}
Laser cooling is a useful technique for the preparation of ultra-cold atoms for use in quantum experiments. The original scheme for Doppler cooling describes an atomic gas driven by a laser that is tuned to be slightly below an electronic transition in the atoms of the gas \cite{hansch,wineland,ashkin}. Due to the Doppler effect, atoms will absorb more photons if they move toward the light source, so they’ll always absorb more photons that oppose their direction of motion. The re-emission of these absorbed photons is isotropic with no net contribution to the atom momentum. The first successful Doppler cooling experiments were reported soon after for Mg$^+$ and Be$^+$ \cite{drullinger,toschek}.

Since the inception of Doppler cooling, other types of cooling schemes have been demonstrated. The Zeeman slower achieved temperatures near the Doppler limit using neutral atoms \cite{philips}. Later, atoms would be cooled below the Doppler limit, the possibility of which was attributed to additional atomic states and laser polarization \cite{chu}. Sub-Doppler temperatures would also be achieved with Sisyphus cooling, and Doppler cooling techniques would later be further refined using optical pumping \cite{dalibard,claude}.

\begin{figure}[h]
  \includegraphics[width=\linewidth]{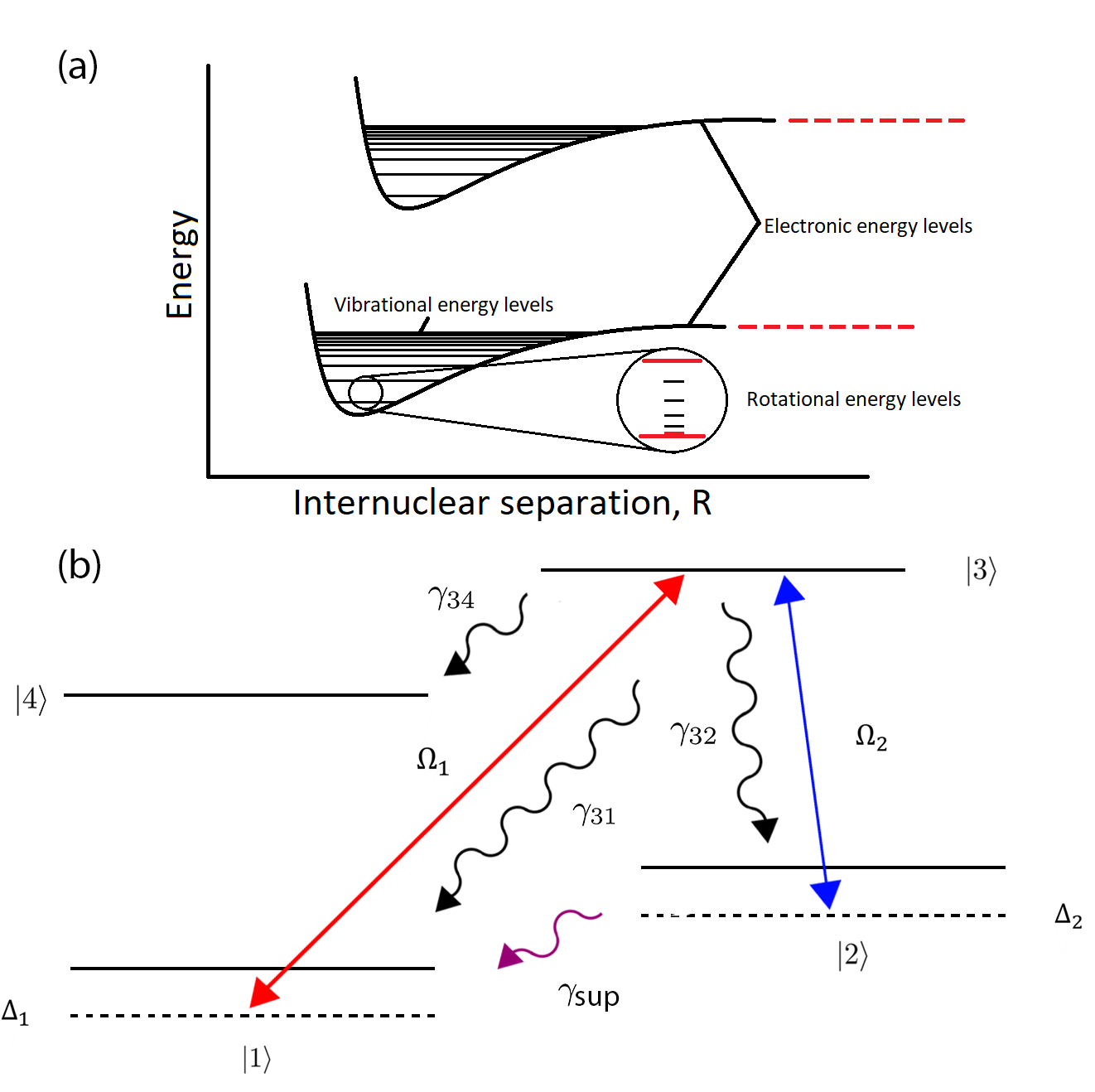}
  \caption{(a) The energy level structure of a molecule. In red is the effective three-level scheme used in our proposal. (b) The $\Lambda$-configuration considered in this study. The $|1\rangle,|2\rangle,\text{ and }|3\rangle$ states are the $\Lambda$ part of the scheme, with modification $|4\rangle$ added to simulate decay into some non-cyclic energy level (we assume $\gamma_{34}\gg\gamma_{31},\gamma_{32},\gamma_{\rm sup}$). Straight arrows represent coupling between the counter-propagating beams and the two main electronic transitions. Only the coupling for two beams are shown here; in the complete model there is a pair of beams for each of six possible directions in space. Squiggly arrows represent spontaneous decay, except for $\gamma_{\rm sup}$, which is the superradiant decay. $|1\rangle$ and $|2\rangle$ are in the same electronic ground state, but with different vibrational modes.}
  \label{fig:lambda}
\end{figure}

The logical next step in the refinement of laser cooling is a method that works for molecules. The fundamental problem with the extension of existing laser cooling techniques to more complex structures is that they rely on cyclic (electronic) transitions. Typical cooling schemes for neutral atoms and ions assume a perfect two-level structure. Ideal structures like this are rare in most molecules. The molecular energy level structure consists of electronic levels that are split by different vibrational and rotational levels (see \cref{fig:lambda}a). Each of these transitions  would require an individual laser to repump the molecule.

Cooling and laser slowing of several oxides and other molecules such as SrF, YO, CaF, YbF, and CaOH has favorable Frank-Condon factors and thus maintain near-cyclic transitions, such that a small number of lasers for repumping are sufficient \cite{shuman,ybf,yeo,lev,stuhl,ding,xu,hummon,hemmerling,caoh}. Cooling of potassium-rubidium molecules to the ultra-cold regime has been demonstrated, and there is a proposed method for cooling polyatomic molecules of six or greater atoms using an attached metal atom as a photon cycling site, for which there also exists near-cyclic transitions \cite{marco,doyle}. Proposals exist for an adiabatic cooling technique that feature two-level transfers, and these proposals are less reliant on spontaneous emission, making them good candidates for cooling of particles without cyclic transitions, such as molecules \cite{norcia,bartolotta}. However, these schemes don't achieve the ultra-cold temperatures that atoms can reach and work almost exclusively for diatomic molecules. Currently there are no other proposals for laser cooling a general class of polyatomic molecules. 

We suggest here to use the two lowest vibrational states of the ground electronic states as a viable cyclic transition in order to cool such a complex system to the ultra-cold regime.

Obviously, using vibrational states for radiative transitions poses two major problems. One is that vibrational radiative lifetimes are forbiddingly long (often of the order of seconds or longer), while electronic excited states of molecules typically have similar lifetimes as atoms. Another is that the photons from a vibrational transition have less energy and thus less momentum than the higher energy photons of an electronic transition. A third, minor, problem, is that vibrational transitions don't have dipole selection rules, so rotational transitions need to be involved.

Superradiance could mitigate the first of these problems. In a superradiant system, molecules cooperatively emit spontaneous photons much faster than they would in vacuum, which is due to the dipoles of the molecules becoming locked in phase \cite{dicke,gross,lin}. The long wavelength of vibrational transitions leads to a very high optical depth, which increases the likelihood of superradiance, and superradiance has been demonstrated to speed up vibrational transitions in atoms and polar molecules \cite{guin}. Recent studies have shown that superradiance can occur on some transitions that are useful in laser cooling schemes \cite{guinthree}. If the vibrational cooling transition, in addition, is pumped using counter-propagating Raman beams, this two-photon transition would impart two high-frequency photon momenta to the molecule,  consequently speeding up the cooling \cite{wang05}. In order to solve the third problem, the rotational transition selection rules can be used. In total, this technique is an ideal candidate for Doppler cooling polyatomic molecules.

The aim of this paper is to describe a novel technique that uses superradiance to laser cool molecules. First the general theory for two- and three-level systems will be reviewed, followed by a detailed explanation incorporating this theory into a four-level scheme in three dimensions. Finally, the results of these simulations will be interpreted, with some concluding discussion regarding the implications and possible expansion of this work.

\section{Model}
We start by providing a short review of laser cooling on a two-level transition. Consider an electronic transition coupled to a beam with Rabi frequency $\Omega$ and detuning $\Delta=\omega-\omega_0$ for driving frequency $\omega$ and transition frequency $\omega_0$, and spontaneous decay rate $\gamma$. The system moves with velocity $v$ in the positive direction. Assume the beam points in the opposite direction of $v$.

The Hamiltonian in the rotating frame is
\begin{equation}
    \frac{H}{\hbar}=
        \left(
\begin{array}{cc}
  0 & -\Omega/2 \\
  -\Omega/2 & \Delta+ kv\\
\end{array}
\right)
\end{equation}
which leads to the master equation
\begin{equation} \label{master}
    \dot{\rho}=-\frac{i}{\hbar}[H,\rho]+\Gamma
\end{equation}
with the typical radiation damping terms
\begin{equation}
   \Gamma=
        \left(
\begin{array}{cc}
  \gamma\rho_{ee} & -\frac{\gamma}{2}\rho_{ge} \\
  -\frac{\gamma}{2}\rho_{eg} & -\gamma\rho_{ee}\\
\end{array}
\right).
\end{equation}
%\textcolor{blue}{After stimulated emission or absorption of one photon, the molecule's momentum changes by $\pm\hbar k$, respectively.} 
Stimulated absorption of one photon with subsequent emission in a random direction leads to the average particle momentum change of $\hbar k$ against the direction of the stimulating laser. Thus the effective cooling experienced by the system is proportional to the difference of %the two rates of stimulation. 
the rates of stimulation of laser fields propagating in different directions.

After adiabatic elimination\footnote{In adiabatic elimination, the coherences are assumed to adapt very quickly to the equilibrium of the populations, resulting in the dynamics being dominated by the populations. Thus we set $\dot{\rho_{ij}}=0$ for $i\neq j$.}, we find the result
\begin{equation}
    \dot{\rho_{ee}}=-\gamma\rho_{ee}+\frac{\gamma\Omega^2}{\gamma^2+4(\Delta+ kv)^2}(\rho_{gg}-\rho_{ee})
\end{equation}
The term $-\gamma\rho_{ee}$ describes how the excited state population changes with spontaneous decay. The remaining two terms are the rate of stimulated absorption and the rate of stimulated emission, respectively. 
Each time a photon is spontaneously emitted, it kicks the molecule via
\begin{equation*}
   \frac{\hbar k}{m}\gamma\rho_{ee}\cos(\theta)
\end{equation*}
where $\theta$ is the angle between the velocity of the molecule and the velocity of the emitted photon, and $\gamma\rho_{ee}$ is the rate of spontaneous emission. These random momentum kicks do not contribute to the average velocity:
\begin{equation}
    \overline{\frac{\hbar k}{m}\gamma\rho_{ee}\cos(\theta)}=0.
\end{equation}
This technique can be readily generalized to more levels, e.g., a three-level $\Lambda$ scheme, with $|1\rangle$ and $|2\rangle$ denoting the two lowest ground state vibrational states, and a single electronic excited state $|3\rangle$. One Raman beam is coupled to the $|1\rangle\leftrightarrow|3\rangle$ transition with with Rabi frequency $\Omega_1$ and detuning $\Delta_1$. A second beam is coupled to the $|1\rangle\leftrightarrow|3\rangle$ transition with Rabi frequency $\Omega_2$ and detuning $\Delta_2$. Note that a photon orbiting up and down this Raman transition will impart momentum $\hbar k$ from the atom/molecule during the stimulated absorption and take one during the stimulated emission. The Hamiltonian in the rotating frame is
\begin{equation}
\frac{H}{\hbar}=
    \left(
\begin{array}{ccc}
 \Delta_1+ k_1v & 0 & -\Omega_1 \\
 0 & \Delta_2+ k_2v & -\Omega_2 \\
 -\Omega_1 & -\Omega_2 & 0 \\
\end{array}
\right)
\end{equation}

For the explicit form of the damping term $\Gamma$, see the appendix. We use \eq{master} to solve the system. Analogously to the two-level system, we find the contribution of the $|1\rangle\leftrightarrow|3\rangle$ and $|2\rangle\leftrightarrow|3\rangle$ transitions to the cooling are
\begin{equation*}
   -\hbar k_1 \frac{\gamma_{31}\Omega_1^2}{\gamma_{31}^2+4(\Delta_1+ kv)^2}(\rho_{11}-\rho_{33})
\end{equation*}
and
\begin{equation*}
    -\hbar k_2 \frac{\gamma_{32}\Omega_2^2}{\gamma_{32}^2+4(\Delta_2+ kv)^2}(\rho_{22}-\rho_{33})
\end{equation*}
respectively, where $\gamma_{ij}$ represents the decay $|i\rangle\to |j\rangle$. $k_1$ ($k_2$) is the wavenumber of the Raman beam coupled to the $|1\rangle\leftrightarrow|3\rangle$ ($|1\rangle\leftrightarrow|3\rangle$) transition. If the Raman transition is kept close to two-photon resonance with the vibrational transition -- thus creating an electromagnetically induced transparency (EIT) resonance -- the excited state $|3\rangle$ will not be populated. Thus, in this case, there is no decay into uncontrolled vibrational states.

In practice, the molecule will occasionally be in its excited state. The main reasons for this departure from ideal EIT come from decoherence due to finite linewidths, collisions, and the Doppler effect. (Note that the undesired Doppler shifts in this setup lead to decoherence only relative to the energy difference of the two lower vibrational states and thus contribute relatively weakly.) This excitation leads to decay into some non-cyclic energy level that is not part of the original $\Lambda$ configuration. We account for this by expanding our Hamiltonian to include a fourth state, which represents the population lost during the cooling process. The density matrix expands accordingly. We assume that decay into this fourth energy level does not contribute to the system's cooling via spontaneous emission.

The decay $\gamma_{super}$ describes a superradiant transition between the two ground states, and it is the principle advantage of this scheme that enables it to work for molecules. $\gamma_{super}$ is large enough to ensure rapid transitions from $|2\rangle\leftrightarrow|1\rangle$, allowing for swift cooling despite the relatively low energy of photons emitted by this transition. The superradiance decay is approximately proportional to the decay between vibration states and the optical depth for the respective transition \cite{lin2012superradiance,ma2022superradiance,rastogi2022superradiance,PhysRevLett.131.033605,od1,od2,od3,od4,od5,od6,od7,od8}:
\begin{equation} \label{sup}
    \gamma_{\rm super}\approx\gamma_{2\rightarrow 1}\times OD.
\end{equation}
The optical depth is approximately proportional to $n\lambda^2 r\rho_{22}$, for population of the upper vibrational state $\rho_{22}$, number density $n$, vibrational transition wavelength $\lambda$, and cloud size of radius $r$. Note that $\gamma_{2\rightarrow 1}$ is the ordinary spontaneous decay rate for the vibrational transition, which is very slow. For superradiance to be strong, the molecules must be separated by a distance less than $\lambda$.

We assume each Raman beam to be Gaussian.
With two pairs of beams oriented along each axis and counter-propagating, it is possible in principal to cool a system with arbitrary velocity in three dimensions. The reason for counter-propagating beams is so that the momentum kick from stimulated absorption and emission add positively.

In this scheme, the ultimate limit (i.e., the case of vanishing dephasing) to Doppler cooling depends only on the superradiant $|2\rangle\rightarrow|1\rangle$ transition. It is
\begin{equation} \label{limit}
    k_BT_D=\hbar\gamma_{super}/2.
\end{equation}

It should be noted that the three-level scheme presented here is actually unrealistic, because all three transitions are assumed to be strongly electric dipole allowed which, because of parity, is impossible. This can be remedied, however, by using a six-level scheme with two subsequent rotational levels, each with opposite parity, for each vibrational level. This additional modification to the original scheme does not change the calculation significantly, and it is analogous for atomic cooling techniques. See the appendix for the more realistic scheme. However, for the purposes of our simulations, we used the scheme presented in \fig{fig:lambda}.

The power requirements for the Raman lasers in our scheme are the same as those of the early molecule cooling techniques \cite{ospelkaus1,ospelkaus2,ospelkaus3,ospelkaus4}.

\section{Implementation}
We describe a four-level system (see \fig{fig:lambda}) with two electronic ground states that differ by vibrational energy and a third electronic excited state. This system is coupled to two pairs of counter propagating Raman beams along each axis, for a total of 12 beams.

In practice, most molecules will occasionally decay into a non-cyclic energy state. We simulate this by adding a fourth state that acts as a loss channel with a strong decay rate $\gamma_{34}$.

A problem of this nature involving coupling of two beams to a system can be solved by working in the rotating frame of the beams. In this scenario there are six pairs of beams interacting with the system simultaneously. Working in the rotating frame of one pair of beams yields time-dependent terms in the Hamiltonian that result from the other pairs. This system cannot be solved in the steady state, so we describe the system with six distinct Hamiltonians (and density matrices), one in the rotating frame of each pair of beams.

In accordance with the approximation, the Hamiltonian that describes interaction in the $\pm$ x-direction is
\begin{equation}
\frac{H^{(\pm x)}}{\hbar}=
    \left(
\begin{array}{cccc}
 \Delta_1\pm k_1v_x & 0 & -\Omega_1 &0 \\
 0 & \Delta_2\pm k_2v_x & -\Omega_2 &0\\
 -\Omega_1 & -\Omega_2 & 0 &0\\
 0&0&0&0\\
\end{array}
\right)
\end{equation}
where $v_x$ is the $x-$component of the velocity. We define $v_x$ to be in the positive direction. An analogous definition is used for $H^{(\pm y)}$ and $H^{(\pm z)}$.

Just as $H^{(\pm x)}$ is the Hamiltonian for interaction in the $\pm$ x-direction, $\rho^{(\pm x)}$ represents the density matrix for interaction in the $\pm$ x-direction.

For a transition $|i\rangle\leftrightarrow|3\rangle$, the average deceleration experienced by the molecule due to stimulated absorption and stimulated emission in the $\pm$ x-direction is
\begin{equation*}
    \pm\frac{1}{m}\hbar k_i\frac{\gamma_{3i}\Omega_i^2}{\gamma_{3i}^2+4(\Delta_i\pm k_iv_x)^2}(\rho^{(\pm x)}_{ii}-\rho^{(\pm x)}_{33}).
\end{equation*}
Finally, we discuss spontaneous decay. For each transition $|3\rangle\to|2\rangle$ and $|3\rangle\to|1\rangle$ there is spontaneous emission. Each time a photon is absorbed by the molecule via one of these transitions, in the x-, y-, or z-directions, it is emitted, but in a random direction, averaging to zero. Thus, the deceleration experienced by the molecule from spontaneous emission in all six directions is
\begin{equation*}
\sum_{i=1,2}\sum_{j=\pm}\sum_{k=x,y,z}\frac{1}{m}\hbar k_i \gamma_{3i}\rho_{33}^{(j,k)}\cos(\theta_k).
\end{equation*}

Numerical simulations of these dynamics lead to \Cref{fig:cooling1,fig:cooling2,fig:cooling3,fig:cooling4,fig:cooling5,fig:cooling6,fig:chart}. For each spatial direction, two Hamiltonians and two density matrices are created, and two equations of motion are written. Then the following actions are performed, in order, for step $i$:
\begin{enumerate}
    \item All six independent equations of motion are solved numerically in the range $[t_i,t_i+\Delta t]$ for timestep $\Delta t$.
    \item The contribution to cooling due to stimulated absorption and emission via each laser is calculated for timestep $\Delta t$. The contribution is computed in each spatial direction independently.
    \item The contribution to cooling from random kicks via spontaneous emission is calculated for $\Delta t$. The average number of spontaneous emissions $\gamma_{3j} \Delta t$ ($j=1,2$) in the range $[t_i,t_i+\Delta t]$ from each spatial direction is calculated independently, then the momentum kick from each emission is applied to the molecule in a random direction.
\end{enumerate}
Each simulation is run for a total time of $400\tau$.

The two transitions in the three-level system are approximated to have the same wavelength, $\lambda = 2\pi/k=$ 500 nm, i.e. the order of magnitude of Rb MOTs at room temperature. Each molecule is approximated to have a mass $m\approx 10^{-27}$ kg. The molecule's initial velocity is given in terms of the recoil velocity, $v_r=\hbar k/m$. The initial velocity was $v_0=$ 75$v_r\approx $ 500 m/s. All other parameters are given in terms of $\gamma=1/\tau$, i.e. the order of a typical optical decay. For all simulations, $\tau = 100$ ns. The various decays in this scheme are as follows:
\[
\gamma_{31}=\gamma_{32}=\gamma_{opt}=\gamma,\text{ }\gamma_{34}=10\gamma,
\]
where $\gamma_{\rm opt}$ is the decay rate on the optical electronic transitions.

To determine the superradiance decay, we return to \cref{sup}. The wavelength for vibrational transitions is typically a few micrometers to a millimeter, density can reach up to $10^9$ cm$^{-3}$, and $r$ is typically on the order of millimeters to centimeters \cite{lin2012superradiance}. Assuming a typical vibrational lifetime of about a second, this gives an $OD$ on the order of $10^0-10^7$. Using \cref{sup}, this leads to
\begin{equation*}
    \gamma_{\rm sup}\approx \gamma\rho_{22}.
\end{equation*}
Simulations were run with Rabi frequencies $\Omega_i$ varying from $0$ to $500\gamma$. Detunings $\Delta_i$ varied from $0$ to $-80\gamma$. The system  was allowed to be two-photon resonant, and $\Delta_1=\Delta_2=\Delta$ and $\Omega_1=\Omega_2=\Omega$ for all simulations except those which produced \fig{fig:chart} . All simulations were run 100 times and their average was computed as the result.

The width of the Raman beams is negligible compared to the width created by the Doppler limit, so we let it equal 0. This is customary for the density matrix approach we have taken.

To compute the theoretical Doppler cooling limit, we make the approximation $\rho_{22}\approx1/2$. With the above parameters and \cref{limit}, we calculate a Doppler cooling limit of $T_D\approx35$ $\mu$K.

\section{Results and Discussion}
\begin{figure}[ht] 
\begin{center}
\includegraphics[width=0.9\linewidth]{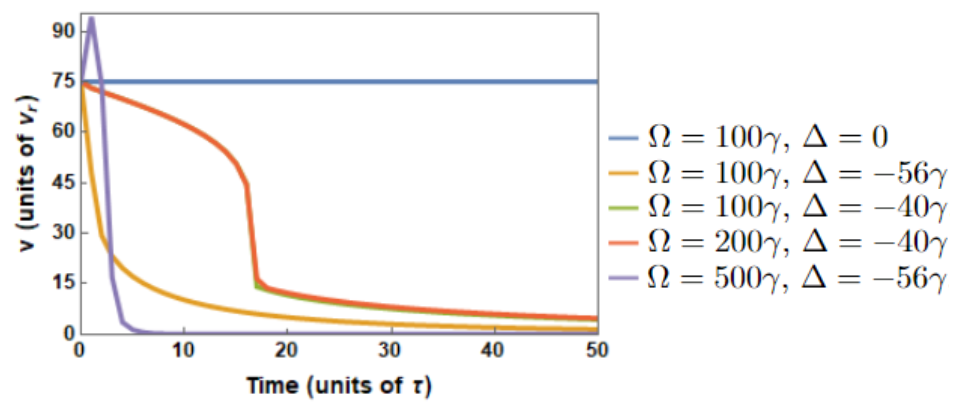}
\end{center}
 \caption{Cooling results for various choices of detuning and Rabi frequency, averaged over 100 simulations. These curves demonstrate the strength of our cooling method for choices of detuning near $-45\gamma$.}
 \label{fig:cooling1}
\end{figure}
\begin{figure}[ht]
\begin{center}
\includegraphics[width=0.9\linewidth]{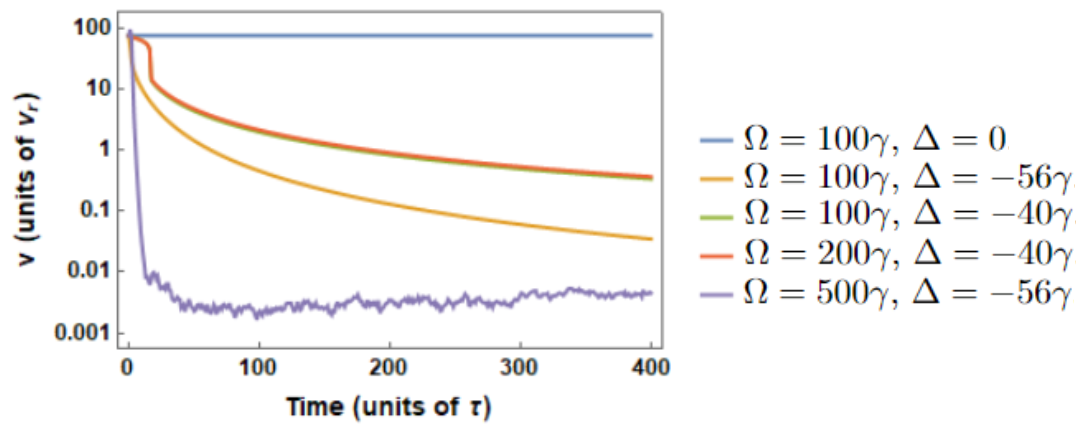}
\end{center}
\caption{Log plot of cooling using various choices of detuning and Rabi frequency, averaged over 100 simulations. In the best cases, the slowest final velocities are achieved in under $50\tau$ and are as low as $10^{-3}v_r$, corresponding to a temperature of about $10^{-9}$K, well below our calculated Doppler limit of $70$ $\mu$K.}
\label{fig:cooling2}
\end{figure}

\begin{figure}[hb]
\begin{center}
\includegraphics[width=0.9\linewidth]{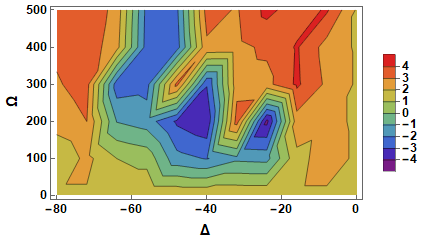}
\end{center}
\caption{Logarithmic plot of molecule velocity after $400\tau$ of cooling, averaged over 100 simulations. Velocity is in units of $v_r$. There is a region where the best cooling occurs. This region is roughly centered on, and split in half by, $\Delta=-45\gamma$}.
\label{fig:cooling3}
\end{figure}

\begin{figure}[ht]
\begin{center}
\includegraphics[width=0.9\linewidth]{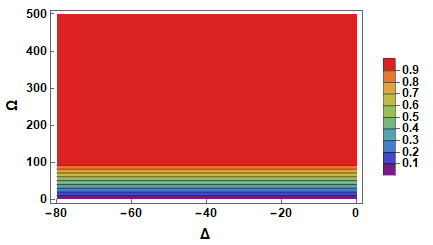}
\end{center}
\caption{Fraction of molecules lost after $400\tau$ of cooling, averaged over 100 simulations. The choice of detuning has very little effect on loss compared to Rabi frequency.}
\label{fig:cooling4}
\end{figure}

\begin{figure}[ht]
\begin{center}
\includegraphics[width=0.9\linewidth]{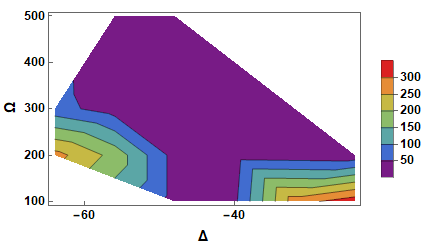}
\end{center}
\caption{Time (in units of $\tau$) to 1 percent cooling, averaged over 100 simulations. Blank space indicates that 1 percent cooling did not occur for that pair of Rabi frequency and detuning. Within the region of good cooling, the molecule can be cooled completely in as few as $50\tau$}.
\label{fig:cooling5}
\end{figure}

\begin{figure}[ht]
\begin{center}
\includegraphics[width=0.9\linewidth]{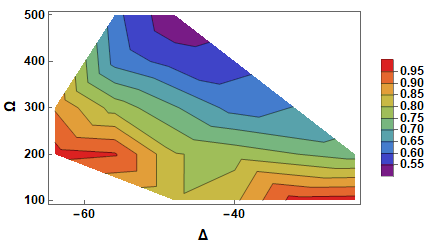}
\end{center}
\caption{Fraction of molecules lost at the instant of 1 percent cooling, averaged over 100 simulations. In the regions corresponding to complete cooling in under $50\tau$, the fraction of population lost can be as low as 55 percent in the best cases.}
\label{fig:cooling6}
\end{figure}

 
An ideal cooling scheme wants to see fast cooling with low loss of population to non-cyclic energy levels. We consider the fraction of final molecular velocity to initial velocity as the percent of cooling. Our simulations show that for proper choice of $\Omega$, cooling of 99 percent or greater can be achieved within a time $10\tau$. Within the parameter regime we examined, the choice of $\Omega$ that gives the most options for choice of $\Delta$ that will result in satisfactory cooling is in the $100\gamma$ to $400\gamma$ range. As shown in \fig{fig:cooling1}a, for $\Omega=200\gamma$, various values of $\Delta$ can be chosen to result in quick cooling, with greater absolute $\Delta$ generally resulting in faster cooling. Final velocity for these simulations ranges from the order of $10^{-1}v_r\text{ to }10^{-3}v_r$ for the best cooling (see \fig{fig:cooling2}). For a complete examination of the parameter regime, see \fig{fig:cooling3}. There is a region inside which cooling of 1 percent or smaller is achieved, with final velocities on the order of blue{$10^{-3}v_r\approx 6.67\times10^{-3}$ m/s and a temperature of $\approx1.6\times 10^{-9}$K} in the best cases. As $\Omega$ increases, the range of ideal choices of $\Delta$ increases. Figures \nofig{fig:cooling4} and \nofig{fig:cooling5} examine the time (in units of $\tau$) and fraction of molecules lost to noncyclical decay, respectively, at the moment of cooling of 1 percent or smaller. Since lost population is roughly exponential with time, the fastest cooling scenarios result in the least population lost. Figure \nofig{fig:cooling6} shows the population lost at the end of cooling. The proportion of population lost depends more strongly on the choice of $\Omega$ compared to $\Delta$.

In general, more rapid cooling is expected for stronger red-detuning. When both single-photon transitions are red-detuned but the two-photon transition is blue-detuned, it should still be possible to achieve reasonable cooling as long as the two-photon transition is near resonance (small absolute two-photon detuning). In all of our simulations the system was two-photon resonant. 

\begin{figure}[h]
\begin{center}
\includegraphics[width=0.9\linewidth]{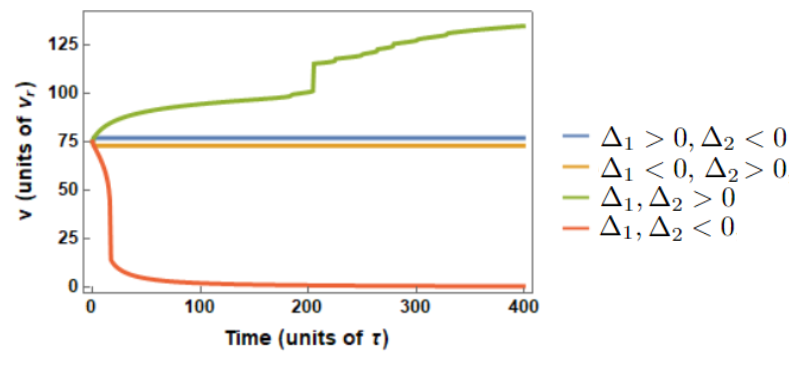}
\end{center}
\caption{Cooling scenarios for the different sign combinations of detunings. For all four scenarios $\Omega=100\gamma$. The sign of $\Delta_1$ dictates whether the molecule will cool or heat up. This is because the first electronic transition scatters more photons than the second electronic transition, which is due to the fact that the upper vibrational state will sometimes decay into the lower vibrational state via superradiant decay. The sign of $\Delta_2$ dictates whether the cooling/heating experienced by the molecule will be fast or slow.}
\label{fig:chart}
\end{figure}

Figure \nofig{fig:chart} shows the results of different sign combinations for the two detunings. Whether heating or cooling occurs is identified by the sign of $\Delta_1$. Whether the heating/cooling is fast or slow is identified by the sign of $\Delta_2$. This is likely because of the fact that the $|3\rangle\to|1\rangle$ transition is faster than the $|3\rangle\to|2\rangle\to|1\rangle$, which contains an extra step.

Consider one of the two transitions. In one dimension, the net force contributing to the cooling is the difference of the cooling forces of the opposing lasers,
\begin{equation*}
   \hbar k \gamma_{opt}\Omega^2(\rho_{11}-\rho_{33})(\frac{1}{\gamma_{opt}^2+4(\Delta+ kv)^2}-\frac{1}{\gamma_{opt}^2+4(\Delta- kv)^2})
\end{equation*}
The value of $\Delta$ that maximizes this expression at $t=0$, or $\rho_{11}=1$, is $\Delta_{max}\approx -45\gamma$. We expect the cooling process to be faster the closer the detunings are to this value. As can be seen in \fig{fig:cooling3}, the best cooling is roughly centered around this detuning value.

In the best cooling cases, the simulated final velocity of the molecules is $\approx6.67\times10^{-3}$ m/s, which corresponds to a temperature of $\approx1.6\times 10^{-9}$K, 4 orders of magnitude below our calculated Doppler limit. It is important to acknowledge that these are the results of semi-classical simulations. To obtain quantitatively reliable results below the limit reached by atomic Doppler cooling, a full quantum treatment taking into account sub-Doppler cooling techniques of our system would be required. The basic interpretation of our results should be that it is possible to cool at least as well as in atomic Doppler cooling, but beyond this our results are quantitatively unreliable. Our scheme should be thought of as the molecular analog to atomic Doppler cooling and follows the typical atomic approach, such as the approach outlined in \cite{foot}, very closely. This is also why, although the counterpropagating beams create a standing wave in each axis, molecules experience negligible positional dependence on the cooling force. Most molecule and atom samples are much smaller than the wavelength. 

\begin{figure}[h] 
\begin{center}
\includegraphics[width=0.9\linewidth]{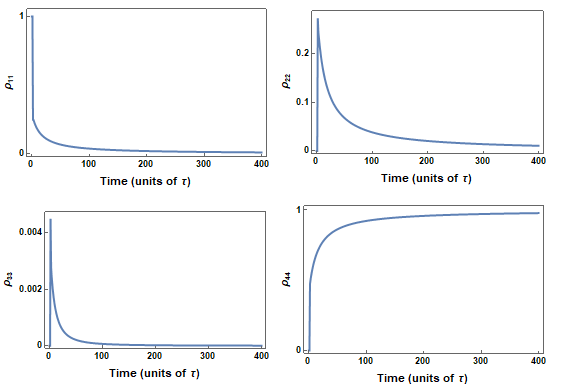}
\end{center}
 \caption{The molecule experiences partial EIT during the cooling process. When cooling is finished, the population in the electronic excited state is close to zero. Most of the population that isn't lost is in one of the two vibrational states, mostly the upper state $|2\rangle$. For $\Omega=100\gamma$ and $\Delta=-56\gamma$, the average populations are pictured.}
 \label{fig:eit}
\end{figure}

One might compare our scheme to velocity selective coherent population trapping (VSCPT) because the molecule does experience some electromagnetically induced transparency (EIT), as shown in \fig{fig:eit}. However, besides this similarity, our scheme has nothing to do with VSCPT. In VSCPT there is no driving force to cool the system. Instead, the system is allowed to relax into the cooling process.

In principle, our scheme still works with the inclusion of hyperfine states. To a first order approximation, they will create additional selection rules and new potential candidates for dark states. Of course, these are issues shared by most atomic cooling techniques.

\begin{table}[h]
\begin{tabular}{|l|l|l|}
\hline
 & $\Delta_1<0$ &  $\Delta_1>0$ \\ \hline
$\Delta_2<0$ & rapid cooling & gradual heating \\ \hline
$\Delta_2>0$ & gradual cooling &  rapid heating \\ \hline
\end{tabular}
\caption{Different cooling scenarios for the four-level scheme by sign combination of the detunings.}
\end{table}

%%\onecolumngrid

%\twocolumngrid

\section{Conclusion}
This paper presents a novel scheme for doppler cooling of molecules where the cycling transition happens along a vibrational transition instead of an electronic one.
%. Previously, doppler schemes used only electronic modes for their state configuration, but finding cyclic transitions of only electronic modes is unrealistic for molecules and other systems more complicated than atoms. 
Our scheme's configuration contains two vibrational states in the electronic ground state. The level scheme presented here is realistic for most molecules if one can choose the lowest two vibrational states. The vibrational transitions' characteristic slow scattering time is alleviated by superradiance, the low photon momentum by a Raman transition in the visible part of the spectrum.
This lends this protocol a unique advantage that allows for even relatively complicated molecules to be cooled with relative efficiency. The seven-level modification to the original scheme accounts for the fact that vibrational transitions don't have selection rules without affecting the calculation presented here in any meaningful way.

Our results indicate that with the proper choice of parameters rapid cooling can be achieved on a system with arbitrary initial state within the studied manifold. Further study of this proposal may benefit from the incorporation of chirping into the Doppler cooling scheme, which was not investigated here. This scheme should be generalized to include more levels and possibly different types of level schemes that describe more realistic molecules. The concept of superradiance-assisted cooling may also be applied to different types of laser cooling altogether. 

\section{Acknowledgments}
We acknowledge funding from the NSF for this work.

\section{Appendix}
The damping term in the four-level scheme is
\begin{equation*}
\begin{split}
\Gamma_{11}&=\gamma_{\rm sup} \rho_{22}+\gamma_{31} \rho_{33},\\
\Gamma_{12}&=-\frac{1}{2} \gamma_{\rm sup} \rho_{12},\\
\Gamma_{13}&=-\frac{1}{2} (\gamma_{31}+\gamma_{32}+\gamma_{34}) \rho_{13},\\
\Gamma_{21}&=-\frac{1}{2} \gamma_{\rm sup} \rho_{21},\\
\Gamma_{22}&=\gamma_{32} \rho_{33}-\gamma_{\rm sup} \rho_{22},\\
\Gamma_{23}&=-\frac{1}{2} \rho_{23} (\gamma_{\rm sup}+\gamma_{31}+\gamma_{32}+\gamma_{34}),\\
\Gamma_{31}&=-\frac{1}{2} (\gamma_{31}+\gamma_{32}+\gamma_{34}) \rho_{31},\\
\Gamma_{32}&=-\frac{1}{2} \rho_{32} (\gamma_{\rm sup}+\gamma_{31}+\gamma_{32}+\gamma_{34}),\\
\Gamma_{33}&=(\gamma_{31}+\gamma_{32}+\gamma_{34}) (-\rho_{33}),\\
\Gamma_{44}&=\gamma_{34}\rho_{33},\\
\Gamma_{4j}&=0,\text{ }j\neq4,\\
\Gamma_{i4}&=0,\text{ }i\neq4.
\end{split}
\end{equation*}

\begin{figure}[H]
\includegraphics[width=1.0\linewidth]{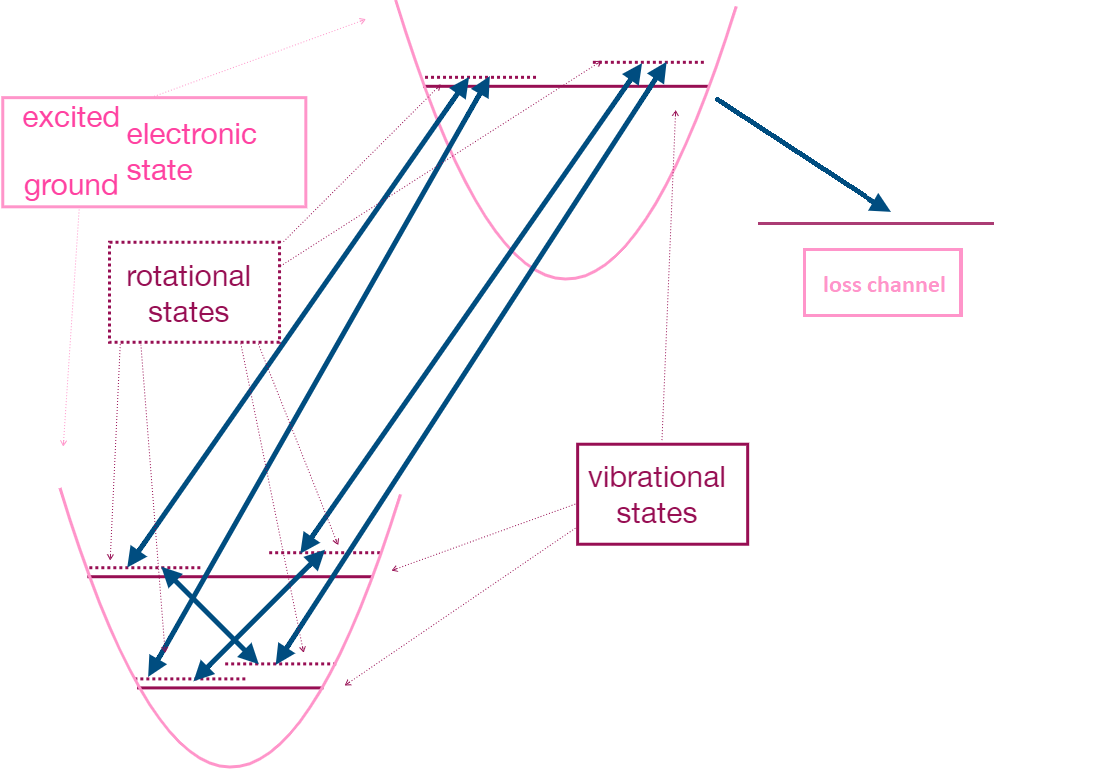}
\caption{The four-level scheme from \fig{fig:lambda}, extended to seven levels to account for parity changes in each transition. The additional levels, which split the original three levels into six, are rotational modes. There is a seventh level, which simulates loss similarly to that of $|4\rangle$ in} \fig{fig:lambda}.
\label{fig:sixlevels}
\end{figure}

The scheme presented in \fig{fig:lambda}b is technically not physical: all three transitions are assumed to be strong electric dipole allowed which, because of parity, is impossible. This can be remedied, however, by using a six-level scheme with two subsequent rotational levels (i.e., with opposite parity) for each vibrational level, like in \fig{fig:sixlevels}. This additional modification to the original scheme does not change the calculation significantly. This is the same as for atomic cooling techniques, which also feature an expansion to account for rotational levels.

\bibliographystyle{IEEEtran}
\bibliography{main}

\end{document}